\documentclass[11pt,english,english,12pt,onecolumn, draftcls]{IEEEtran}
\usepackage[T1]{fontenc}
\usepackage[latin9]{inputenc}
\usepackage{units}
\usepackage{amsthm}
\usepackage{amsmath}
\usepackage{amssymb}
\usepackage{graphicx}
\usepackage{esint}
\PassOptionsToPackage{version=3}{mhchem}
\usepackage{mhchem}

\makeatletter
\theoremstyle{plain}
\newtheorem{thm}{\protect\theoremname}
\theoremstyle{definition}
\newtheorem{defn}[thm]{\protect\definitionname}
\theoremstyle{remark}
\newtheorem{rem}[thm]{\protect\remarkname}
\theoremstyle{plain}
\newtheorem{prop}[thm]{\protect\propositionname}

\ifCLASSINFOpdf
\else
\fi
\theoremstyle{plain}
\newtheorem{propo}{Proposition}
\renewenvironment{prop}{\begin{propo}}{\end{propo}}
\theoremstyle{remark}
\newtheorem{rema}{Remark}
\renewenvironment{rem}{\begin{rema}}{\end{rema}}

\usepackage{babel}

\providecommand{\definitionname}{Definition}
\providecommand{\propositionname}{Proposition}
\providecommand{\remarkname}{Remark}
\providecommand{\theoremname}{Theorem}

\makeatother

\usepackage{babel}
\providecommand{\definitionname}{Definition}
\providecommand{\propositionname}{Proposition}
\providecommand{\remarkname}{Remark}
\providecommand{\theoremname}{Theorem}

\begin{document}

\title{Two-way Communication with Adaptive Data Acquisition }

\author{\IEEEauthorblockN{Behzad Ahmadi and Osvaldo Simeone}\\
 \IEEEauthorblockA{CWCSPR, ECE Dept.\\
 New Jersey Institute of Technology\\
 Newark, NJ, 07102, USA\\
 Email: \{behzad.ahmadi,osvaldo.simeone@njit.edu\}} }

\maketitle

%







\begin{abstract}
Motivated by computer networks and machine-to-machine communication
applications, a bidirectional link is studied in which two nodes,
Node 1 and Node 2, communicate to fulfill generally conflicting informational
requirements. Node 2 is able to acquire information from the environment,
e.g., via access to a remote data base or via sensing. Information
acquisition is expensive in terms of system resources, e.g., time,
bandwidth and energy and thus should be done efficiently by adapting
the acquisition process to the needs of the application. As a result
of the forward communication from Node 1 to Node 2, the latter wishes
to compute some function, such as a suitable average, of the data
available at Node 1 and of the data obtained from the environment.
The forward link is also used by Node 1 to query Node 2 with the aim
of retrieving suitable information from the environment on the backward
link. The problem is formulated in the context of multi-terminal rate-distortion
theory and the optimal trade-off between communication rates, distortions
of the information produced at the two nodes and costs for information
acquisition at Node 2 is derived. The issue of robustness to possible
malfunctioning of the data acquisition process at Node 2 is also investigated.
The results are illustrated via an example that demonstrates the different
roles played by the forward communication, namely data exchange, query
and control.\end{abstract}
\IEEEpeerreviewmaketitle

\section{Introduction}

In computer networks and machine-to-machine links, communication is
often interactive and serves a number of integrated functions, such
as data exchange, query and control. As an exemplifying example, consider
the set-up in Fig. 1 in which the terminals labeled Node 1 and Node
2 communicate on bidirectional links. Node 2 has access to a data
base or, more generally, is able to acquire information from the environment,
e.g., through sensors. As a result of the communication on the forward
link, Node 2 wishes to compute some function, e.g., a suitable average,
of the data available at Node 1 and of the information retrievable
from the environment. Instead, Node 1 queries Node 2 on the forward
link with the aim of retreiving some information from the environment
through the backward link. 

\begin{figure}[h!]
\centering\includegraphics[bb=18bp 8bp 529bp 235bp,clip,scale=0.6]{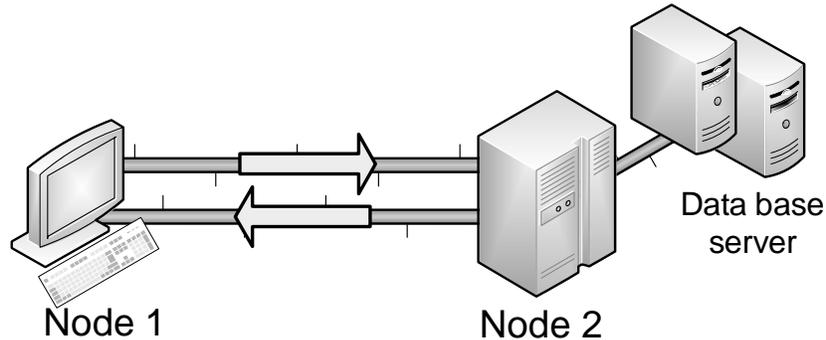}
\caption{Two-way communication with adaptive data acquisition. }

\label{fig:db} 
\end{figure}

Information acquisition from the environment is generally expensive
in terms of system resources, e.g., time, bandwidth or energy. For
instance, accessing a remote data base requires interfacing with a
server by following the appropriate protocol, and activating sensors
entails some energy expenditure. Therefore, data acquisition by Node
2 should be performed efficiently by adapting to the informational
requirements of Node 1 and Node 2. 

To summarize the discussion above, in the system of Fig. 1 the forward
communication from Node 1 to Node 2 serves three integrated purposes:
\emph{i}) \emph{Data exchange}: Node 1 provides Node 2 with the information
necessary for the latter to compute the desired quantities; \emph{ii})
\emph{Query}: Node 1 informs Node 2 about its own informational requirements,
to be met via the backward link; \emph{iii}) \emph{Control}: Node
1 instructs Node 2 on the most effective way to perform data acquisition
from the environment in order to satisfy Node 1's query and to allow
Node 2 to perform the desired computation.

This work sets out to analyze the setting in Fig. 1 from a fundamental
theoretical standpoint via information theory. Specifically, the problem
is formulated within the context of network rate-distortion theory,
and the optimal communication strategy, involving the elements of
data exchange, query and control, is identified. Examples are worked
out to illustrate the relevance of the developed theory. Finally,
the issue of robustness is tackled by assuming that, unbeknownst to
Node 1, Node 2 may be unable to acquire information from the environment,
due, e.g., to energy shortages or malfunctioning. The optimal robust
strategy is derived and the examples extended to account for this
generalized model.

\subsection{Related Work}

The work in this paper builds on the long line of research within
network information theory that deals with source coding with side
information (see, e.g., \cite{Elgammal} for an introduction). More
specifically, we adopt the model of a side information {}``vending
machine\textquotedblright{} that has been introduced in \cite{Permuter}.
This model accounts for source coding scenarios in which acquiring
information at the receiver entails some cost and thus should be done
efficiently. Specifically, in this model, the quality of the side
information $Y$ can be controlled at the decoder by selecting an
action $A$ that affects the effective channel between the source
$X$ and the side information $Y$ through a conditional distribution
$p_{Y|X,A}(y|x,a)$. The distribution $p_{Y|X,A}(y|x,a)$ defines
the side information {}``vending machine'' as per the nomenclature
of \cite{Permuter}. Each action $A$ is associated with a cost, and
the problem is that of characterizing the available trade-offs among
rate, distortion and action cost. We emphasize the conventional formulation
of the source coding problem with side information instead assumes
that the relationship between source and side information is determined
by a given conditional distribution $p_{Y|X}(y|x)$ that cannot be
controlled.

Various works have extended the results in \cite{Permuter}. Extensions
to multi-terminal models can be found in \cite{Weissman_multi}. Specifically,
references \cite{Weissman_multi}-\cite{Kittipong-secure} considered
a set-up analogous to the Heegard-Berger problem \cite{HB,Kaspi},
in which the side information may or may not be available at the decoder.
In \cite{Ahmadi_DSC}, a distributed source coding setting that generalizes
\cite{Berger-Yeung} to the case of a decoder with a side information
{}``vending machine'' is investigated. Multi-hop models were studied
in \cite{Ahmadi_DSC}\cite{Ahmadi_Chiru}. 
\begin{figure}[h!]
\centering\includegraphics[bb=145bp 530bp 522bp 669bp,clip,scale=0.7]{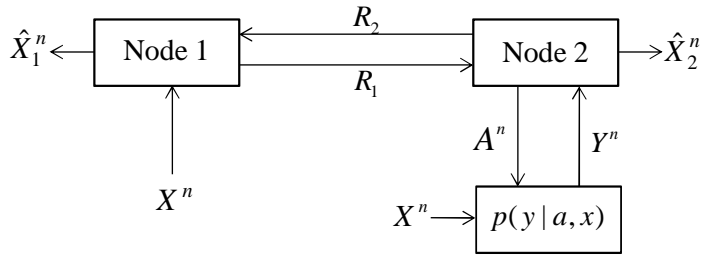}
\caption{Two-way source coding with a side information vending machine at Node
2.}

\label{fig:fig1} 
\end{figure}
In \cite{Compression with actions}, a related problem is considered
in which the sequence to be compressed is dependent on the actions
taken by a separate encoder. Other extensions include \cite{Kittipong,Kittipong-secure}
where the model of \cite{Permuter} is revisited under the additional
constraints of common reconstruction \cite{Steinberg} or of secrecy
with respect to an \textquotedbl{}eavesdropping\textquotedbl{} node.

In this paper, the model of a side information {}``vending machine''
is used to model the information acquisition process at Node 2 in
Fig. \ref{fig:db}. Unlike \cite{Permuter} and the previous work
discussed above, communication between Node 1 and Node 2 is assumed
to be \emph{bidirectional}. The problem of characterizing the rate-distortion
region for a two-way source coding models, with conventional action-independent
side information sequences at Node 2 has been addressed in \cite{Kaspi-2way,Weissman_2way,Ma}
and references therein.

\subsection{Contributions and Organization of the Paper}

This work studies the model in Fig. 1, which is detailed in terms
of a block diagram in Fig. 2. The system model is introduced in Sec.
\ref{sub:System-Model_2way}. The optimal trade-off between the rates
of the bidirectional communication, the distortions of the reconstructions
of the desired quantities at the two nodes, and the budget for information
acquisition at Node 2 is derived in Sec. \ref{sub:RD_2way}. An example
that illustrates the application of the developed theory is discussed
in Sec. \ref{sub:Example}. Finally, in Sec. \ref{sub:2way-HB_ind},
the results are extended to the scenario in Fig. \ref{fig:fig4} in
which, unbeknownst to Node 1, Node 2 may be unable to perform information
acquisition. 

\emph{Notation}: Throughout the paper, a random variable is denoted
by an upper case letter~(e.g., $X,Y,Z$) and its realization is denoted
by a lower case letter~(e.g., $x,y,z$). Moreover, the shorthand
notation $X^{n}$ is used to denote the tuple (or the column vector)
of random variables $(X_{1},\ldots,X_{n})$, and $x^{n}$ is used
to denote a realization. We define $[a,b]=[a,a+1,...,b]$ for $a\leq b$
and $[a,b]=\emptyset$, otherwise. We say that $X\textrm{---}Y\textrm{---}Z$
forms a Markov chain if $p(x,y,z)=p(x)p(y|x)p(z|y)$, that is, if
$X$ and $Z$ are conditionally independent of each other given $Y$.

\section{System Model\label{sub:System-Model_2way}}

The two-way source coding problem of interest, sketched in Fig. \ref{fig:fig1},
is formally defined by the probability mass functions (pmfs) $p_{X}(x)$
and $p_{Y|AX}(y|a,x)$, and by the discrete alphabets $\mathcal{X},\mathcal{Y},\mathcal{A},\mathcal{\hat{X}}_{1},\mathcal{\hat{X}}_{2},$
along with distortion and cost metrics to be discussed below. The
source sequence $X^{n}=(X_{1},...,X_{n})\in\mathcal{X}^{n}$ consists
of  $n$ independent and identically distributed (i.i.d.) entries
$X_{i}$ for $i\in[1,n]$ with pmf $p_{X}(x)$. Node 1 measures sequence
$X^{n}$ and encodes it in a message $M_{1}$ of $nR_{1}$ bits, which
is delivered to Node 2. Node 2 wishes to estimate a sequence $\hat{X}_{2}^{n}\in\mathcal{\hat{X}}_{2}^{n}$
within given distortion requirements. To this end, Node 2 receives
message $M_{1}$ and based on this, it selects an action sequence
$A^{n},$ where $A^{n}\in\mathcal{A}^{n}.$ 

The action sequence affects the quality of the measurement $Y^{n}$
of sequence $X^{n}$ obtained at the Node 2. Specifically, given $A^{n}$
and $X^{n}$, the sequence $Y^{n}$ is distributed as $p(y^{n}|a^{n},x^{n})=\prod_{i=1}^{n}p_{Y|A,X}(y_{i}|a_{i},x_{i})$.
The cost of the action sequence is defined by a cost function $\Lambda$:
$\mathcal{A\rightarrow}[0,\Lambda_{\max}]$ with $0\leq\Lambda_{\max}<\infty,$
as $\Lambda(a^{n})=\sum_{i=1}^{n}\Lambda(a_{i})$. The estimated sequence
$\hat{X}_{2}^{n}$ with $\hat{X}_{2}^{n}\in\mathcal{\hat{X}}_{2}^{n}$
is then obtained as a function of $M_{1}$ and $Y^{n}$. 

Upon reception on the forward link, Node 2 maps the message $M_{1}$
received from Node 1 and the locally available sequence $Y^{n}$ in
a message $M_{2}$ of $nR_{2}$ bits, which is delivered back to Node
1. Node 1 estimates a sequence $\hat{X}_{1}^{n}\in\mathcal{\hat{X}}_{1}^{n}$
as a function of $M_{2}$ and $X^{n}$ within given distortion requirements. 

The quality of the estimated sequence $\hat{X}_{j}^{n}$ is assessed
in terms of the distortion metrics $d_{j}(x,y,\hat{x}_{j})$: $\mathcal{X}\times\mathcal{Y}\times\mathcal{\hat{X}}_{j}\rightarrow\Bbb{R_{+}}\cup\{\infty\}$
for $j=1,2,$ respectively. Note that this implies that $\hat{X}_{j}^{n}$
is allowed to be a lossy version of any function of the source and
side information sequences. A more general model is studied in Sec.
\ref{sub:RD_2way-ind}. It is assumed that $D_{j}=\min_{\hat{x}_{j}\in\hat{\mathcal{X}}_{j}}E[d(X,Y,\hat{X}_{j})]<\infty$
 for $j=1,2$. A formal description of the operations at encoder and
decoder follows. 
\begin{defn}
\label{def_2way}An $(n,R_{1},R_{2},D_{1},D_{2},\Gamma,\epsilon)$
code for the set-up of Fig. \ref{fig:fig1} consists of a source encoder
for Node 1 
\begin{equation}
\mathrm{g}_{1}\text{: }\mathcal{X}^{n}\rightarrow[1,2^{nR_{1}}],\label{encoder1}
\end{equation}
which maps the sequence $X^{n}$ into a message $M_{1};$ an {}``action\textquotedblright{}\ function
\begin{equation}
\mathrm{\ell}\text{: }[1,2^{nR_{1}}]\times\mathcal{Y}^{i-1}\rightarrow\mathcal{A},\label{action_fun}
\end{equation}
which maps the message $M_{1}$ and the previously observed into an
action sequence $A^{n};$ a source encoder for Node 2 
\begin{equation}
\mathrm{g}_{2}\text{:}\text{ }\mathcal{Y}^{n}\times[1,2^{nR_{1}}]\rightarrow[1,2^{nR_{2}}],\label{encoder2}
\end{equation}
which maps the sequence $Y^{n}$ and message $M_{1}$ into a message
$M_{2};$ two decoders, namely 
\begin{equation}
\mathrm{h}_{1}\text{: }[1,2^{nR_{2}}]\times\mathcal{X}^{n}\rightarrow\mathcal{\hat{X}}_{1}^{n},\label{decoder1}
\end{equation}
which maps the message $M_{2}$ and the sequence $X^{n}$ into the
estimated sequence $\hat{X}_{1}^{n};$ 
\begin{equation}
\mathrm{h}_{2}\text{: }[1,2^{nR_{1}}]\times\mathcal{Y}^{n}\rightarrow\mathcal{\hat{X}}_{2}^{n},\label{decoder2}
\end{equation}
which maps the message $M_{1}$ and the sequence $Y^{n}$ into the
estimated sequence $\hat{X}_{2}^{n};$ such that the action cost constraint
$\Gamma$ and distortion constraints $D_{j}$ for $j=1,2$ are satisfied,
i.e., 
\begin{align}
\frac{1}{n}\underset{i=1}{\overset{n}{\sum}}\mathrm{E}\left[\Lambda(A_{i})\right] & \leq\Gamma\label{action cost}\\
\text{ and }\frac{1}{n}\underset{i=1}{\overset{n}{\sum}}\mathrm{E}\left[d_{j}(X_{i},Y_{i},\hat{X}_{ji})\right] & \leq D_{j}\text{ for }j=1,2.\label{dist const}
\end{align}

\end{defn}

\begin{defn}
\label{def_ach}Given a distortion-cost tuple $(D_{1},D_{2},\Gamma)$,
a rate tuple $(R_{1},R_{2})$ is said to be achievable if, for any
$\epsilon>0$, and sufficiently large $n$, there exists a $(n,R_{1},R_{2},D_{1}+\epsilon,D_{2}+\epsilon,\Gamma+\epsilon)$
code.
\end{defn}

\begin{defn}
\label{def_reg}The \textit{rate-distortion-cost region }$\mathcal{R}(D_{1},D_{2},\Gamma)$
is defined as the closure of all rate tuples $(R_{1},R_{2})$ that
are achievable given the distortion-cost tuple $(D_{1},D_{2},\Gamma)$. \end{defn}
\begin{rem}
For the special case in which the side information $Y$ independent
of the action $A$ given $X$, i.e., for $\ p(y|a,x)=p(y|x),$ the
rate-distortion region $\mathcal{R}(D_{1},D_{2},\Gamma)$ has been
derived in \cite{Kaspi-2way}. Instead, if $D_{2}=D_{2,max}$, the
set of all achievable rates $R_{1}$ was characterized in \cite{Permuter}.

\begin{rem}The definition (\ref{action_fun}) of an action encoder
allows for adaptation of the actions to the previously observed values
of the side information $Y$. This possibility was studied in \cite{Chiru}
for the point-to-point one-way model, which is obtained by setting
$R_{2}=0$ in the setting of Fig. \ref{fig:fig1}. \end{rem}
\end{rem}
In the following sections, for simplicity of notation, we drop the
subscripts from the definition of the pmfs, thus identifying a pmf
by its argument.

\section{Rate-Distortion-Cost Region \label{sub:RD_2way}}

In this section, a single-letter characterization of the rate-distortion-cost
region is derived. 
\begin{prop}
\label{prop:RD_action_2way}The rate-distortion-cost region $\mathcal{R\mbox{\ensuremath{(D_{1},D_{2},\Gamma)}}}$
for the two-way source coding problem illustrated in Fig. \ref{fig:fig1}
is given by the union of all rate pairs $(R_{1},R_{2})$ that satisfy
the conditions\begin{subequations}\label{eqn: RD_action_2way} 
\begin{eqnarray}
R_{1} & \geq & I(X;A)+I(X;U|A,Y)\label{eq:R1}\\
\ce{and}\mbox{ }R_{2} & \geq & I(Y;V|A,X,U),\label{eq:R2}
\end{eqnarray}
\end{subequations}where the mutual information terms are evaluated
with respect to the joint pmf 
\begin{align}
p(x,y,a,u,v)=p(x)p(a,u|x)p(y|a,x)p(v|a,u,y) & ,\label{eq:joint}
\end{align}
for some pmfs $p(a,u|x)$ and $p(v|a,u,y)$ such that the inequalities\begin{subequations}\label{eqn: action_const}
\begin{eqnarray}
\ce{E}[d_{1}(X,Y,\ce{f}_{1}(V,X))] & \leq & D_{1},\label{eq:dist1}\\
\ce{E}[d_{2}(X,Y,\ce{f}_{2}(U,Y))] & \leq & D_{2},\label{eq:dist2}\\
\ce{and}\textrm{ }\ce{E}[\Lambda(A)] & \leq & \Gamma,\label{eq:action_bound}
\end{eqnarray}
\end{subequations}are satisfied for some function $\ce{f}_{1}\textrm{: }\mathcal{V}\times\mathcal{X}\rightarrow\hat{\mathcal{X}}_{1}$
and $\ce{f}_{2}\textrm{: }\mathcal{U}\times\mathcal{Y}\rightarrow\hat{\mathcal{X}}_{2}$.
Finally, $U$ and $V$ are auxiliary random variables whose alphabet
cardinality can be constrained as $|\mathcal{U}|\leq|\mathcal{X}||\mathcal{A}|+4$
and $|\mathcal{V}|\leq|\mathcal{U}||\mathcal{Y}||\mathcal{A}|+1$
without loss of optimality. \end{prop}
\begin{rem}
For the special case in which the side information $Y$ is independent
of the action $A$ given $X$$,$ i.e., for $p(y|a,x)=p(y|x),$ the
rate-distortion region $\mathcal{R}(D_{1},D_{2},\Gamma)$ in Proposition
\ref{prop:RD_action_2way} reduces to that derived in \cite{Kaspi-2way,Weissman_2way}.
Instead, if $D_{2}=D_{2,max}$, the result reduces to that in \cite{Permuter}.
\end{rem}
The proof of the converse is provided in Appendix A. The achievability
follows as a combination of the techniques proposed in \cite{Permuter}
and \cite{Kaspi-2way}, and requires the forward link to be used,
in an integrated manner, for data exchange, query and control. Specifically,
for the forward link, similar to \cite{Permuter}, Node 1 uses a successive
refinement codebook. Accordingly, the base layer is used by Node 1
to instruct Node 2 on which actions are best tailored to fulfill the
informational requirements of both Node 1 and Node 2. This base layer
thus represents control information that also serves the purpose of
querying Node 2 in view of the backward communication. We observe
that Node 1 selects this base layer as a function of the source $X^{n}$,
thus allowing Node 2 to adapt its actions for information acquisition
to the current realization of the source $X^{n}$. The refinement
layer of the code used by Node 1 is leveraged, instead, to provide
additional information to Node 2 in order to meet Node 2's distortion
requirement. Node 2 then employs standard Wyner-Ziv coding (i.e.,
binning) \cite{Elgammal} for the backward link to satisfy Node 1's
distortion requirement.

We now briefly outline the main technical aspects of the achievability
proof, since the details follow from standard arguments and do not
require further elaboration here. To be more precise, Node 1 first
maps sequence $X^{n}$ into the action sequence $A^{n}$ using the
standard joint typicality criterion. This mapping requires a codebook
of rate $I(X;A)$ (see, e.g., \cite[pp. 62-63]{Elgammal})$.$ Given
the sequence $A^{n}$, the description of sequence $X^{n}$ is further
refined through mapping to a sequence $U^{n}$. This requires a codebook
of size $I(X;U|A,Y)$ for each action sequence $A^{n}$ using Wyner-Ziv
binning with respect to side information $Y^{n}$ \cite[pp. 62-63]{Elgammal}.
In the reverse link, Node 2 employs Wyner-Ziv coding for the sequence
$Y^{n}$ by leveraging the side information $X^{n}$ available at
Node 1 and conditioned on the sequences $U^{n}$ and $A^{n}$, which
are known to both Node 1 and Node 2 as a result of the communication
on the forward link. This requires a rate equal to the right-hand
side of (\ref{eq:R2}). Finally, Node 1 and Node 2 produce the estimates
$\hat{X}_{1}^{n}$ and $\hat{X}_{2}^{n}$ as the symbol-by-symbol
functions $\hat{X}_{1i}=\textrm{f}_{1}(V_{i},X_{i})$ and $\hat{X}_{2i}=\textrm{f}_{2}(U_{i},Y_{i})$
for $i\in[1,n]$, respectively.
\begin{rem}
The achievability scheme discussed above uses actions that do not
adapt to the previous values of the side information $Y$. The fact
that this scheme attains the optimal performance characterized in
Proposition \ref{prop:RD_action_2way} shows that, as demonstrated
in \cite{Chiru} for the one-way model with $R_{2}=0$, adaptive actions
do not improve the rate-distortion performance.
\end{rem}

\subsection{Indirect Rate-Distortion-Cost Region \label{sub:RD_2way-ind}}

\begin{figure}[h!]
\centering\includegraphics[bb=88bp 530bp 536bp 669bp,clip,scale=0.65]{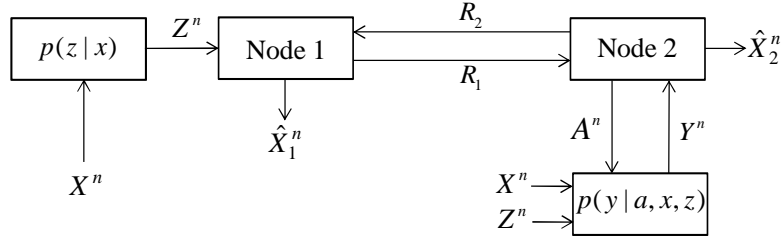}
\caption{Indirect two-way source coding with a side information vending machine
at Node 2.}

\label{fig:fig2} 
\end{figure}

In this section, we consider a more general model in which Node 1
observes only a noisy version of the source $X^{n}$, as depicted
in Fig. \ref{fig:fig2}. Following \cite{Witsenhausen}, we refer
to this setting as posing an indirect source coding problem. The example
studied in Sec. \ref{sub:Example} illustrates the relevance of this
generalization. The system model is as defined in Sec. \ref{sub:System-Model_2way}
with the following differences. The source encoder for Node 1
 
\begin{equation}
\mathrm{g}_{1}\text{: }\mathcal{Z}^{n}\rightarrow[1,2^{nR_{1}}],\label{encoder1_ind}
\end{equation}
maps the sequence $Z^{n}$ into a message $M_{1};$ the decoder for
Node 1 
\begin{equation}
\mathrm{h}_{1}\text{: }[1,2^{nR_{2}}]\times\mathcal{Z}^{n}\rightarrow\mathcal{\hat{X}}_{1}^{n},\label{decoder1_ind}
\end{equation}
maps the message $M_{2}$ and the sequence $Z^{n}$ into the estimated
sequence $\hat{X}_{1}^{n};$ given $(X^{n},A^{n},Z^{n})$, the side
information $Y^{n}$ is distributed as $p(y^{n}|a^{n},x^{n},z^{n})=\prod_{i=1}^{n}p_{Y|A,X,Z}(y_{i}|a_{i},x_{i},z_{i})$
and the distortion constraints are given as 
\begin{align}
\frac{1}{n}\underset{i=1}{\overset{n}{\sum}}\mathrm{E}\left[d_{j}(X_{i},Y_{i},Z_{i},\hat{X}_{ji})\right] & \leq D_{j}\text{ for }j=1,2,\label{dist const_ind}
\end{align}
for some distortion metrics $d_{j}(x,y,z,\hat{x}_{j}):$$\mathcal{X}\times\mathcal{Y}\times{\cal Z}\times\mathcal{\hat{X}}_{j}\rightarrow\Bbb{R_{+}}\cup\{\infty\}$,
for $j=1,2$. The next proposition derives a single-letter characterization
of the rate-distortion-cost region. 
\begin{prop}
\label{prop:RD_action_2way_ind}The rate-distortion-cost region $\mathcal{R\mbox{\ensuremath{(D_{1},D_{2},\Gamma)}}}$
for the indirect two-way source coding problem illustrated in Fig.
\ref{fig:fig2} is given by the union of all rate pairs $(R_{1},R_{2})$
that satisfy the conditions\begin{subequations}\label{eqn: RD_action_2way_ind}
\begin{eqnarray}
R_{1} & \geq & I(Z;A)+I(Z;U|A,Y)\label{eq:R1_ind}\\
\ce{and}\mbox{ }R_{2} & \geq & I(Y;V|A,Z,U),\label{eq:R2_ind}
\end{eqnarray}
\end{subequations}where the mutual information terms are evaluated
with respect to the joint pmf 
\begin{align}
p(x,y,z,a,u,v)\negmedspace=\negmedspace p(x,z)p(a,\negmedspace u|z)p(y|a,\negmedspace x,\negmedspace z)p(v|a,\negmedspace u,\negmedspace y),\label{eq:joint-ind}
\end{align}
for some pmfs $p(a,u|x)$ and $p(v|a,u,y)$ such that the inequalities\begin{subequations}\label{eqn: action_const-ind}
\begin{eqnarray}
\ce{E}[d_{1}(X,Y,Z,\ce{f}_{1}(V,Z))] & \leq & D_{1},\label{eq:dist1_ind}\\
\ce{E}[d_{2}(X,Y,Z,\ce{f}_{2}(U,Y))] & \leq & D_{2},\label{eq:dist2-1}\\
\ce{and}\textrm{ }\ce{E}[\Lambda(A)] & \leq & \Gamma,\label{eq:action_bound-ind}
\end{eqnarray}
\end{subequations}are satisfied for some function $\ce{f}_{1}\textrm{: }\mathcal{V}\times\mathcal{Z}\rightarrow\hat{\mathcal{X}}_{1}$
and $\ce{f}_{2}\textrm{: }\mathcal{U}\times\mathcal{Y}\rightarrow\hat{\mathcal{X}}_{2}$.
Finally, $U$ and $V$ are auxiliary random variables whose alphabet
cardinality can be constrained as $|\mathcal{U}|\leq|\mathcal{Z}||\mathcal{A}|+3$
and $|\mathcal{V}|\leq|\mathcal{U}||\mathcal{Y}||\mathcal{A}|+1$
without loss of optimality. 
\end{prop}
The proof of the achievability and converse follows with slight modifications
from that of Proposition \ref{prop:RD_action_2way}. Specifically,
in the achievability the sequence $X^{n}$ is replaced by its noisy
version, i.e., the sequence $Z^{n}$, and the rest of the proof remains
essentially unchanged. The proof of the converse is provided in Appendix
A.

\section{Example\label{sub:Example}}

In this section, we consider a binary example for the set-up in Fig.
\ref{fig:fig2} to illustrate the main aspects of the problem and
the relevance of the theoretical results derived above. Specifically,
we assume binary alphabets as ${\cal X}={\cal A}=\{0,1\}$ and a source
distribution $X\sim\textrm{Bern}(0.5)$. Moreover, the source $Z^{n}$
measured by Node 1 is an erased version of the source $X^{n}$ with
erasure probability $\epsilon$. This means that $Z_{i}=\textrm{e}$,
where $\textrm{e}$ represents an erasure, with probability $\epsilon$
and $Z_{i}=X_{i}$ with probability $1-\epsilon$, for $i\in[1,n]$. 

The vending machine at Node 2 operates as follows:
\begin{eqnarray}
Y & = & \left\{ \begin{array}{lc}
X & \mbox{ for }A=1\\
\phi & \textrm{\mbox{ for }}A=0
\end{array},\right.\label{eq:VM}
\end{eqnarray}
with cost constraint $\Lambda(a)=a$, for $a\in\{0,1\},$ where $\phi$
is a dummy symbol representing the case in which no useful information
is acquired by Node 2. This model implies that a cost budget of $\Gamma$
limits the average number of samples of the sequence $Y$ that can
be measured by Node 2 to around $n\Gamma$ given the constraint (\ref{action cost}).

Node 1 wishes to reconstruct a lossy version of the source $X^{n}$,
while Node 2 is interested in $Z^{n}$. The distortion functions are
the Hamming metrics $d_{1}(x,\hat{x}_{1})=1_{\{x\neq\hat{x}_{1}\}}$
and $d_{2}(z,\hat{x}_{2})=1_{\{z\neq\hat{x}_{2}\}}$. To obtain analytical
insight into the rate-distortion-cost region, in the following we
focus on a number of special cases.

\subsection{$D_{1}=D_{1,max}$ and $D_{2}=0$\label{sub:ex1}}

Consider the distortion requirements $D_{1}=D_{1,max}$ and $D_{2}=0$.
As a result, Node 1 requires no backward communication from Node 2,
while Node 2 wishes to recover $Z^{n}$ losslessly. For the given
distortions, the rate-cost region in Proposition \ref{prop:RD_action_2way_ind}
can be evaluated as \begin{subequations}\label{eqn: r_ex1} 
\begin{eqnarray}
R_{1} & \geq & H_{2}(\epsilon)+(1-\epsilon-\Gamma)^{+}\label{eq:R1_ex1}\\
\ce{and}\mbox{ }R_{2} & \geq & 0,\label{eq:R2_ex1}
\end{eqnarray}
\end{subequations}for any cost budget $\Gamma\geq0$, where $H_{2}(\alpha)=-\alpha\ce{log_{2}}\alpha-(1-\alpha)\ce{log_{2}}(1-\alpha)$
is the binary entropy function. 

A formal proof of this result can be found in Appendix B. The rate
region (\ref{eqn: r_ex1}) shows that, as the cost budget $\Gamma$
for information acquisition increases, the required rate $R_{1}$
decreases down to the rate $H_{2}(\epsilon)$ that is required to
describe only the erasures process $E^{n}$ with $E_{i}=1_{\{Z_{i}=\ce{e}\}}$,
$i=1,...,n$, losslessly to Node 2. This can be explained by noting
that the following time-sharing strategy achieves region (\ref{eqn: r_ex1})
and is thus optimal. 

Node 1 describes the process $E^{n}$ losslessly to Node 2 with $H_{2}(\epsilon)$
bits per symbol. In order to obtain a lossless reconstruction of $Z^{n}$,
Node 2 needs to be informed about $Z_{i}=X_{i}$ for all $i$ in which
$E_{i}=0$. This information can be interpreted as control data that
is used by Node 2 to adapt its information acquisition process. Note
that we have around $n(1-\epsilon)$ such samples of $Z_{i}$. Node
1 describes $Z_{i}=X_{i}$ for $n(1-\epsilon-\Gamma)^{+}$ of these
samples, while the remaining $n\ce{min}(\Gamma,1-\epsilon)$ are measured
by Node 2 through the vending machine. An alternative strategy based
directly on Proposition \ref{prop:RD_action_2way_ind} can be found
in Appendix B. 

Fig. \ref{fig:plot1} illustrates the rate $R_{1}$ in (\ref{eq:R1_ex1})
versus the cost budget $\Gamma$ for $\epsilon=0.2$. We observe that
if $\Gamma\geq1-\epsilon=0.8$ no further improvement of the rate
is possible as per (\ref{eq:R1_ex1}).

\subsection{$D_{1}=0$ and $D_{2}=D_{2,max}$\label{sub:ex2}}

Here we consider the dual case in which Node 1 wishes to reconstruct
sequence $X^{n}$ losslessly ($D_{1}=0$), while Node 2 does not have
any distortion requirements ($D_{2}=D_{2,max}$). As shown in Appendix
B, if $\Gamma\geq\epsilon$, the rate-cost region is given by the
union of all rate pairs $(R_{1},R_{2})$ such that\begin{subequations}\label{eqn: r_ex2}
\begin{eqnarray}
R_{1} & \geq & H_{2}(\epsilon)-\Gamma H\Bigl(\frac{\epsilon}{\Gamma}\Bigr)\label{eq:R1_ex2}\\
\ce{and}\mbox{ }R_{2} & \geq & \epsilon.\label{eq:R2_ex2}
\end{eqnarray}
\end{subequations}Moreover, for $\Gamma<\epsilon$, the region is
empty as the lossless reconstruction of $X$ at Node 1 is not feasible. 

A proof of this result based on Proposition \ref{prop:RD_action_2way_ind}
can be found in Appendix B. In the following, we argue that a natural
time-sharing strategy, akin to that used for the case $D_{1}=D_{1,max},\textrm{}D_{2}=0$
above, would be suboptimal, implying that the optimal strategy requires
a more sophisticated approach based on the successive refinement code
presented in Sec. \ref{sub:RD_2way}.

A natural time-sharing strategy would be the following. Node 1 describes
$n\eta$ samples of the erasure process $E^{n}$, for some $0\leq\eta\leq1$,
losslessly to Node 2, using rate $R_{1}=\eta H_{2}(\epsilon)$. This
information is used by Node 1 to query Node 2 about the desired information.
Specifically, Node 2 sets $A_{i}=1$ if $E_{i}=1$, thus observing
around $n\eta\epsilon$ samples $Y_{i}=X_{i}$ from the vending machine.
These samples are needed to fulfill the distortion requirements of
Node 1. For all the remaining $n(1-\eta$) samples, for which Node
2 does not have control information from Node 1, Node 2 sets $A_{i}=1$,
thus acquiring all the side information samples. Again, this is necessary
given Node 1's requirements. Node 2 conveys losslessly the $n\eta\epsilon$
samples $Y_{i}=X_{i}$ obtained when $E_{i}=1$, which requires $\eta\epsilon$
bits per sample, along with the $n(1-\eta)$ samples $Y_{i}$ in the
second set, which amount instead to $(1-\eta)H(X|Z)$ bits per sample.
Note that we have the rate $H(X|Z)$ by the Slepian-Wolf theorem \cite[Chapter 10]{Elgammal},
since Node 1 has side information $Z_{i}$ for the second set of samples.
Overall, we have $R_{2}=\eta\epsilon+(1-\eta)\epsilon=\epsilon$ bits/source
symbol. This entails a cost budget of $\Gamma=\eta\epsilon+1-\eta$,
and thus $\eta=\nicefrac{(1-\Gamma)}{(1-\epsilon)}$. 

\begin{figure}[h!]
\centering\includegraphics[bb=35bp 300bp 575bp 731bp,clip,scale=0.6]{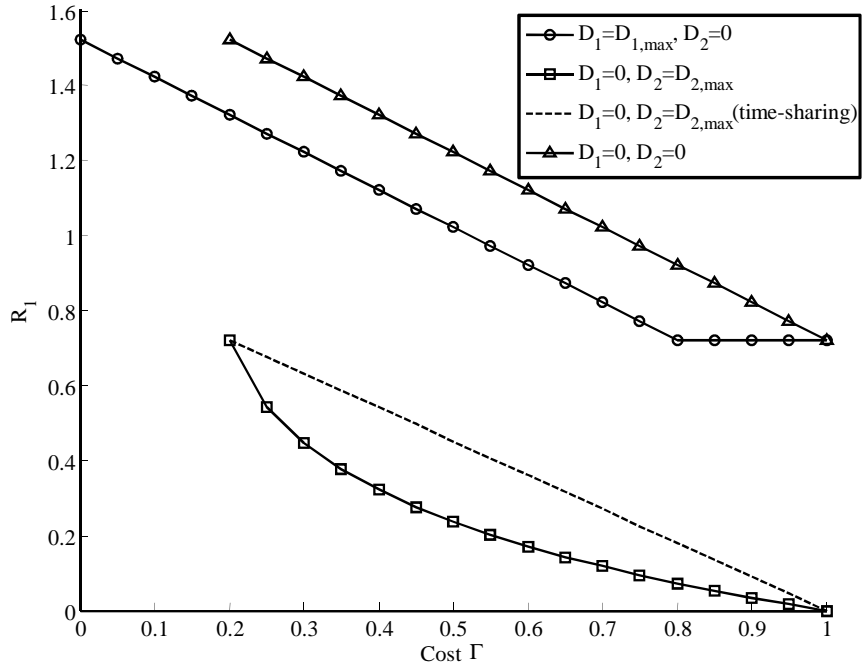}
\caption{Rate $R_{1}$ versus cost $\Gamma$ for the examples in Sec. \ref{sub:Example}
with $\epsilon=0.2$.}

\label{fig:plot1} 
\end{figure}

Fig. \ref{fig:plot1} compares the rate $R_{1}$ as in (\ref{eq:R1_ex2})
with the corresponding rate obtained via time-sharing, for $\epsilon=0.2$.
As seen, in this second case the time-sharing strategy is strictly
suboptimal.

\subsection{$D_{1}=D_{2}=0$\label{sub:ex3}}

We now consider the case in which both nodes wish to achieve lossless
reconstruction, i.e., $D_{1}=D_{2}=0$. As seen in the previous case,
achieving $D_{1}=0$ is not possible if $\Gamma<\epsilon$ and thus
this is a fortiori true for $D_{1}=D_{2}=0$. For $\Gamma\geq\epsilon$,
the rate-cost region is given by\begin{subequations}\label{eqn: r_ex3}
\begin{eqnarray}
R_{1} & \geq & H_{2}(\epsilon)+(1-\Gamma)\label{eq:R1_ex3}\\
\ce{and}\mbox{ }R_{2} & \geq & \epsilon,\label{eq:R2_ex3}
\end{eqnarray}
\end{subequations}as shown in Appendix B. 

A time-sharing strategy that achieves (\ref{eqn: r_ex3}) is as follows.
Node 1 describes the process $E^{n}$ losslessly to Node 2 with $H_{2}(\epsilon)$
bits per symbol. This information serves the functions of query and
control for Node 2. In order to satisfy its distortion requirement,
Node 2 now needs to be informed about $Z_{i}=X_{i}$ for all $i$
in which $E_{i}=0$. Note that we have $n(1-\epsilon)$ such samples
of $Z_{i}$. Node 1 describes $Z_{i}=X_{i}$ for $n(1-\Gamma)\leq n(1-\epsilon)$
of these samples, while the remaining $n(\Gamma-\epsilon)$ are measured
by Node 2 through the vending machine. Node 2 compresses losslessly
the sequence of around $n\epsilon$ samples of $X_{i}$ with $i$
such that $E_{i}=1$ which requires $R_{2}=\epsilon$ bits per sample. 

Fig. \ref{fig:plot1} illustrates the rate $R_{1}$ in (\ref{eq:R1_ex3})
versus the cost budget $\Gamma$ for $\epsilon=0.2$.

\section{When the Side Information May Be Absent\label{sub:2way-HB_ind}}

In this section, we generalize the results of the previous section
to the scenario in Fig. \ref{fig:fig4} in which, unbeknownst to Node
1, Node 2 may be unable to perform information acquisition due, e.g.,
to energy shortage or malfunctioning. This set-up is illustrated in
Fig. \ref{fig:fig4}.

\subsection{System Model\label{sub:System-Model_2way-HB}}

\begin{figure}[h!]
\centering\includegraphics[bb=71bp 530bp 520bp 733bp,clip,scale=0.7]{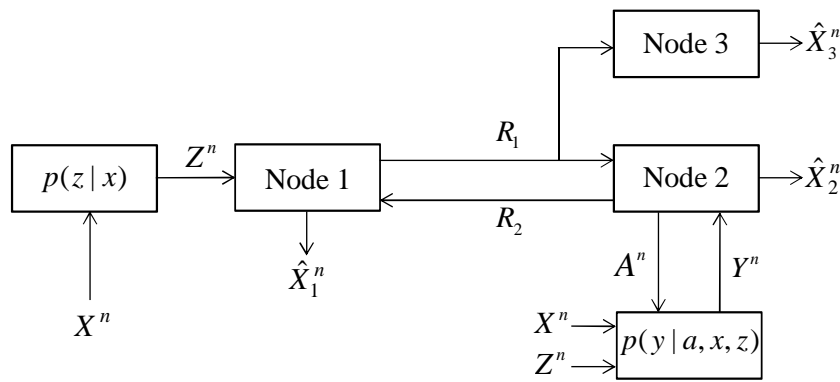}
\caption{Indirect two-way source coding when the side information vending machine
may be absent at the recipient of the message from Node 1.}

\label{fig:fig4} 
\end{figure}

The formal description of an $(n,R_{1},R_{2},D_{1},D_{2},D_{3},\Gamma,\epsilon)$
code for the set-up of Fig. \ref{fig:fig4} is given as in Sec. \ref{sub:RD_2way-ind}
(which generalizes the model in Sec. \ref{sub:System-Model_2way})
with the addition of Node 3. This added node, which has no access
to the side information, models the situation in which the recipient
of the communication from Node 1 happens to be unable to acquire information
from the environment. Note that the same message $M_{1}$ from Node
1 is received by both Node 2 and Node 3. This captures the fact that
the information about whether or not the recipient is able to access
the side information is not available to Node 1. The model in Fig.
\ref{fig:fig4} is a generalization of the so called Heegard-Berger
problem \cite{HB,Kaspi}.

Formally, Node 3 is defined by the decoding function
\begin{equation}
\mathrm{h}_{3}\text{: }[1,2^{nR_{1}}]\rightarrow\mathcal{\hat{X}}_{3}^{n},\label{decoder3}
\end{equation}
which maps the message $M_{1}$ into the the estimated sequence $\hat{X}_{3}^{n};$
and the additional distortion constraint 
\begin{align}
\frac{1}{n}\underset{i=1}{\overset{n}{\sum}}\mathrm{E}\left[d_{3}(X_{i},Y_{i},Z_{i},\hat{X}_{3i})\right] & \leq D_{3}.\label{dist const-HB}
\end{align}
We remark that adding a link between Node 3 and Node 1 cannot improve
the system performance given that Node 3 has only available the message
$M_{1}$ received from Node 1. Therefore, this link is not included
in the model.

\subsection{Rate-Distortion-Cost Region \label{sub:RD_2way-HB_ind}}

In this section, a single-letter characterization of the rate-distortion-cost
region is derived for the set-up in Fig. \ref{fig:fig4}. 
\begin{prop}
\label{prop:RD_action_2way-HB_ind}The rate-distortion-cost region
$\mathcal{R\mbox{\ensuremath{(D_{1},D_{2},D_{3},\Gamma)}}}$ for the
two-way source coding problem illustrated in Fig. \ref{fig:fig4}
is given by the union of all rate pairs $(R_{1},R_{2})$ that satisfy
the conditions\begin{subequations}\label{eqn: RD_action_2way-HB_ind}
\begin{eqnarray}
R_{1} & \negmedspace\geq\negmedspace & I(Z;A)+I(Z;\hat{X}_{3}|A)\nonumber \\
 &  & +I(Z;U|A,Y,\hat{X}_{3})\label{eq:R1-HB}\\
\ce{and}\mbox{ }R_{2} & \negmedspace\geq\negmedspace & I(Y;V|A,Z,U,\hat{X}_{3}),\label{eq:R2-HB}
\end{eqnarray}
\end{subequations}where the mutual information terms are evaluated
with respect to the joint pmf 
\begin{align}
p(x,y,z,a,u,v)= & p(x,z)p(a,u,\hat{x}_{3}|z)p(y|a,x,z)\nonumber \\
 & p(v|a,u,y,\hat{x}_{3}),\label{eq:joint-HB_ind}
\end{align}
for some pmfs $p(a,u,\hat{x}_{3}|z)$ and $p(v|a,u,y)$ such that
the inequalities\begin{subequations}\label{eqn: action_const-ind-HB-1}
\begin{eqnarray}
\ce{E}[d_{1}(X,Y,Z,\ce{f}_{1}(V,Z))] & \leq & D_{1},\label{eq:dist1_ind-HB}\\
\ce{E}[d_{2}(X,Y,Z,\ce{f}_{2}(U,Y))] & \leq & D_{2},\label{eq:dist2-ind_HB}\\
\ce{E}[d_{3}(X,Y,Z,\hat{X}_{3})] & \leq & D_{3},\\
\ce{and}\textrm{ }\ce{E}[\Lambda(A)] & \leq & \Gamma,\label{eq:action_bound-ind-HB}
\end{eqnarray}
\end{subequations}are satisfied for some function $\ce{f}_{1}\textrm{: }\mathcal{V}\times\mathcal{Z}\rightarrow\hat{\mathcal{X}}_{1}$
and $\ce{f}_{2}\textrm{: }\mathcal{U}\times\mathcal{Y}\rightarrow\hat{\mathcal{X}}_{2}$.
Finally, $U$ and $V$ are auxiliary random variables whose alphabet
cardinality can be constrained as $|\mathcal{U}|\leq|\mathcal{Z}||\mathcal{A}|+3$
and $|\mathcal{V}|\leq|\mathcal{U}||\mathcal{Y}||\mathcal{A}|+1$
without loss of optimality.
\end{prop}
The proof of the converse is provided in Appendix A. The achievable
rate (\ref{eq:R1-HB}) can be interpreted as follows. Node 1 uses
a successive refinement code with three layers. The first layer is
defined as for Sec. \ref{sub:RD_2way} and carries query and control
information. The second and third layers are designed as in the optimal
Heegard-Berger scheme \cite{HB}. Specifically, the second layer is
destined to both Node 2 and Node 3, while the third layer targets
only Node 2, which has enhanced decoding capabilities due to the availability
of side information. 

To provide further details, as for Proposition \ref{prop:RD_action_2way},
the encoder first maps the input sequence $Z^{n}$ into an action
sequence $A^{n}$ so that the two sequences are jointly typical, which
requires $I(Z;A)$ bits/source sample. Then, it maps $Z^{n}$ into
the estimate $\hat{X}_{3}^{n}$ for Node 3 using a conditional codebook
with rate $I(Z;\hat{X}_{3}|A)$. Finally, it maps $Z^{n}$ into another
sequence $U^{n}$ using the fact that Node 2 has the action sequence
$A^{n}$, the estimate $\hat{X}_{3}^{n}$ and the measurement $Y^{n}$.
Using conditional codebooks (with respect to $\hat{X}_{3}^{n}$ and
$A^{n}$) and from the Wyner-Ziv theorem, this requires $I(Z;U|A,Y,\hat{X}_{3})$
bit/source sample. As for the rate (\ref{eq:R2-HB}), Node 2 employs
Wyner-Ziv coding for the sequence $Y^{n}$ by leveraging the side
information $Z^{n}$ available at Node 1 and conditioned on the sequences
$U^{n}$, $A^{n}$ and $\hat{X}_{3}^{n}$, which are known to both
Node 1 and Node 2 as a result of the forward communication. This requires
a rate equal to the right-hand side of (\ref{eq:R2-HB}). Finally,
Node 1 and Node 2 produce the estimates $\hat{X}_{1}^{n}$ and $\hat{X}_{2}^{n}$
as the symbol-by-symbol functions $\hat{X}_{1i}=\textrm{f}_{1}(V_{i},Z_{i})$
and $\hat{X}_{2i}=\textrm{f}_{2}(U_{i},Y_{i})$ for $i\in[1,n]$,
respectively.

\subsection{Example\label{sub:Example-HB}}

In this section, we extend the binary example of Sec. \ref{sub:Example}
to the set-up in Fig. \ref{fig:fig4}. Specifically, we consider the
same setting as in Sec. \ref{sub:Example}, with the addition of Node
3. For the latter, we assume a ternary reconstruction alphabet ${\cal \hat{X}}_{3}=\{0,1,*\}$
and the distortion metric $d_{3}(x,z,\hat{x}_{3})=d_{3}(\mathbf{1}_{\{Z=\ce{e}\}},\hat{x}_{3})$
in Table \ref{table2}, where we recall that $E_{i}=\mathbf{1}_{\{Z_{i}=\ce{e}\}}$
is the erasure process. Accordingly, Node 3 is interested in recovering
the erasure process $E^{n}$ under an erasure distortion metric (see,
e.g., \cite{Verdu}) , where {}``{*}'' represents the {}``don't
care'' or erasure reproduction symbol 
\begin{table}
\centering\includegraphics[bb=120bp 560bp 415bp 650bp,clip,scale=0.7]{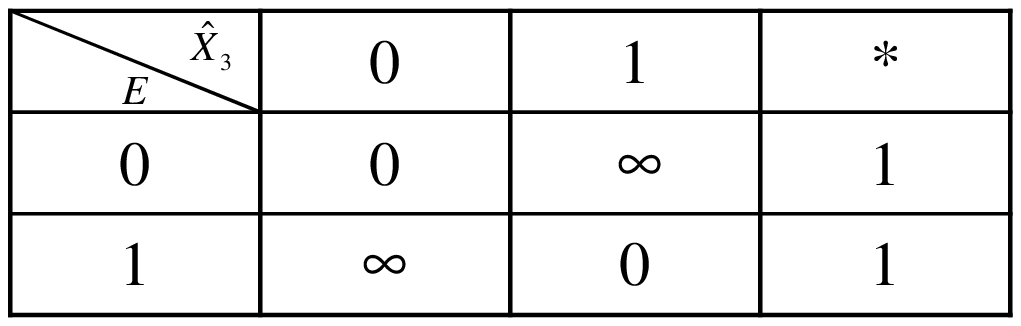}\caption{Erasure distortion for reconstruction at Node 3.}

\label{table2}
\end{table}

We first observe that for cases 1) and 3) in Sec. \ref{sub:Example}
the distortion requirements of Node 3 do not change the rate-distortion
function. This is because, as discussed in Sec. \ref{sub:Example},
the requirement that $D_{2}$ be equal to zero entails that the erasure
process $E^{n}$ be communicated losslessly to Node 2 without leveraging
the side information from the vending machine (which cannot provide
information about the erasure process). It follows that one can achieve
$D_{3}=0$ at no additional rate cost. We thus now focus on the case
2) in Sec. \ref{sub:Example}, namely $D_{1}=0$ and $D_{2}=D_{2,max}$.

In the case at hand, Node 1 wishes to recover $X^{n}$ losslessly,
Node 2 has no distortion requirements and Node 3 wants to recover
$E^{n}$ with distortion $D_{3}$. As explained in Sec. \ref{sub:ex2},
in order to reconstruct $X^{n}$ losslessly at Node 1 we must have
$\Gamma\geq\epsilon$ and $\ce{Pr}(A=1|Z=\ce{e})=1$. Moreover, due
to symmerty of the problem with respect to $Z=0$ and $Z=1$, we can
set  $\ce{Pr}(A=1|Z=0)=\ce{Pr}(A=1|Z=1)=\frac{\gamma-\epsilon}{1-\epsilon}$,
for some $0\leq\gamma\leq\Gamma$. To evaluate the rate-distortion-cost
region (\ref{eqn: RD_action_2way-HB_ind}), we then define $\ce{Pr}(\hat{X}_{3}=*|A=1,Z=\ce{e})\overset{\triangle}{=}p_{1}$,
$\ce{Pr}(\hat{X}_{3}=*|A=0,Z=0)\overset{\triangle}{=}p_{2}$ and $\ce{Pr}(\hat{X}_{3}=*|A=1,Z=0)\overset{\triangle}{=}p_{3}$.
We thus get that the rate-distortion-cost region is given by\begin{subequations}\label{eqn: r_ex3-HB}

\begin{align}
R_{1} & \geq H_{2}(\epsilon)+1-\epsilon-(1-\Gamma)(1-p_{2})-(\Gamma-\epsilon)\nonumber \\
 & (1-p_{3})-(1-\Gamma)p_{2}-\Bigl(\epsilon p_{1}+(\Gamma-\epsilon)p_{3}\Bigr)\nonumber \\
 & \Bigl(H_{2}(\frac{\epsilon p_{1}}{\epsilon p_{1}+(\Gamma-\epsilon)p_{3}})+\frac{(\Gamma-\epsilon)p_{3}}{\epsilon p_{1}+(\Gamma-\epsilon)p_{3}}\Bigr)\label{eq:R1_ex2_HB}\\
\ce{and}\mbox{ }R_{2} & \geq\epsilon,\label{eq:R2_ex2_HB}
\end{align}
\end{subequations}where parameters $p_{1},p_{2},p_{3}\in[0,1]$ must
be selected so as to satisfy the distortion constraint of Node 3,
namely $D_{3}\geq\epsilon p_{1}+(1-\Gamma)p_{2}+(\Gamma-\epsilon)p_{3}$.

\begin{figure}[h!]
\centering\includegraphics[bb=35bp 170bp 575bp 600bp,clip,scale=0.6]{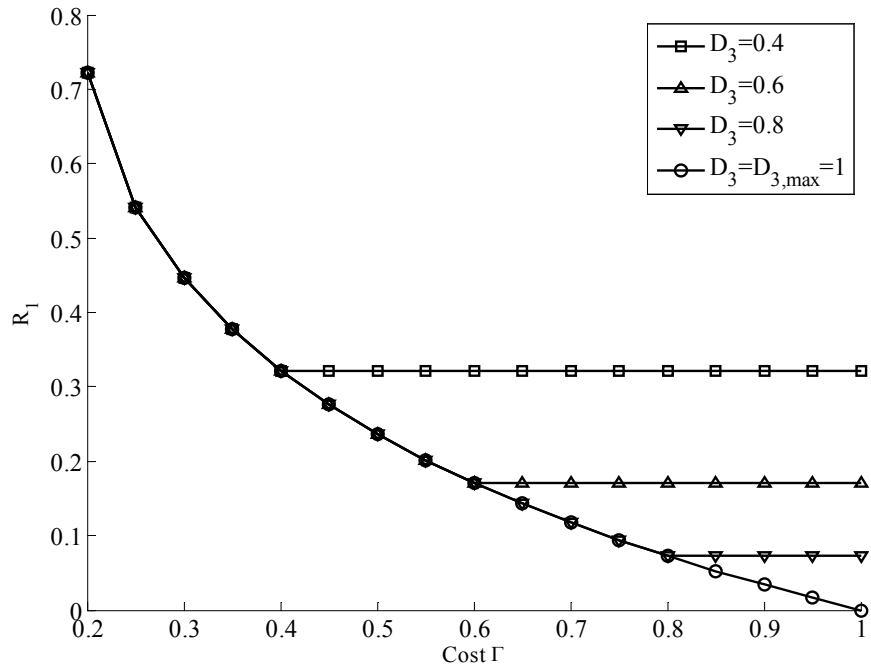}
\caption{Rate $R_{1}$ versus cost $\Gamma$ for the examples in Sec. \ref{sub:Example-HB}
with $\epsilon=0.2$, $D_{1}=0$ and $D_{2}=D_{2,max}$.}

\label{fig:plot2} 
\end{figure}

Fig. \ref{fig:plot2} illustrates the rate $R_{1}$ in (\ref{eq:R1_ex2_HB}),
minimized over $p_{1},p_{2}$ and $p_{3}$ under the constraints mentioned
above versus the cost budget $\Gamma$ for $\epsilon=0.2$ and different
values of $D_{3}$, namely $D_{3}=0.4,$ $0.6$, $0.8$ and $D_{3}=D_{3,max}=1$.
Note that for $D_{3}=D_{3,max}=1$ we obtain the rate in (\ref{eq:R1_ex2}).
As it can be seen, for $\Gamma\leq D_{3}$, the rate decreases with
increasing cost $\Gamma$, but for $\mbox{\ensuremath{\Gamma\geq}}D_{3}$
the rate remains constant while increasing $\Gamma$. The reason is
that for the latter region, i.e., $\Gamma\geq D_{3}$, the performance
of the system is dominated by the distortion requirement of Node 3
and thus increasing the cost budget $\Gamma$ does not improve the
rate. Instead, for $\Gamma\leq D_{3}$, it is sufficient to cater
only to Node 2, and Node 3 is able to recover $E$ with distortion
$D_{3}=\Gamma$ at no additional rate cost.

\section{Concluding Remarks}

For applications such as complex communication networks for cloud
computing or machine-to-machine communication, the bits exchanged
by two parties serve a number of integrated functions, including data
transmission, control and query. In this work, we have considered
a baseline two-way communication scenario that captures some of these
aspects. The problem is addressed from a fundamental theoretical standpoint
using an information theoretic formulation. The analysis reveals the
structure of optimal communication strategies and can be applied to
elaborate on specific examples, as illustrated in the paper. This
work opens a number of possible avenues for future research, including
the analysis of scenarios in which more than one round of interactive
communication is possible \cite{Ma}.

\appendices{ }

\section*{Appendix A: Converse Proof for Proposition \ref{prop:RD_action_2way}}

Here, we prove the converse part of Proposition \ref{prop:RD_action_2way}.
For any $(n,R_{1},R_{2},D_{1}+\epsilon,D_{2}+\epsilon,\Gamma+\epsilon)$
code, we have the series of inequalities
\begin{align}
 & nR_{1}\geq H(M_{1})\nonumber \\
 & \stackrel{(a)}{=}I(M_{1};X^{n},Y^{n})\nonumber \\
 & =H(X^{n})+H(Y^{n}|X^{n})\nonumber \\
 & -H(Y^{n}|M_{1})-H(X^{n}|Y^{n},M_{1})\nonumber \\
 & \stackrel{(b)}{\geq}\sum_{i=1}^{n}H(X_{i})-H(X_{i}|X_{i+1}^{n},Y^{n},M_{1},A^{i})\nonumber \\
 & +H(Y_{i}|Y^{i-1},X^{n},M_{1},A_{i})-H(Y_{i}|Y^{i-1},M_{1},A_{i})\nonumber \\
 & \stackrel{(c)}{\geq}\negmedspace\sum_{i=1}^{n}\negmedspace H\negmedspace(X_{i})\negmedspace-\negmedspace H\negmedspace(X_{i}|A_{i},Y_{i},U_{i})\negmedspace+\negmedspace H\negmedspace(Y_{i}|X_{i},\negmedspace A_{i})\negmedspace-\negmedspace H\negmedspace(Y_{i}|A_{i}),\label{eq::R1_end}
\end{align}
where $(a)$ follows since $M_{1}$ is a function of $X^{n}$ and
since conditioning reduces entropy; $(b)$ follows since $A_{i}$
is a function of $(M_{1},Y^{i-1})$ and $M_{1}$ is a function of
$X^{n}$ and $(c)$ follows since conditioning decreases entropy,
by defining $U_{i}=(M_{1},X_{i+1}^{n},A^{i-1},Y^{i-1})$ and using
the fact that the vending machine is memoryless. We also have the
series of inequalities
\begin{align}
 & nR_{2}\geq H(M_{2})\nonumber \\
 & \geq H(M_{2}|X^{n},M_{1})\nonumber \\
 & \stackrel{(a)}{=}I(M_{2};Y^{n}|X^{n},M_{1})\nonumber \\
 & \stackrel{(b)}{=}\sum_{i=1}^{n}H(Y_{i}|Y^{i-1},X^{n},M_{1},A^{i})\nonumber \\
 & -H(Y_{i}|Y^{i-1},X^{n},M_{1},M_{2},A^{i})\nonumber \\
 & \stackrel{(c)}{\geq}\negmedspace\sum_{i=1}^{n}\negmedspace H\negmedspace(Y_{i}|X_{i},\negmedspace A_{i},\negmedspace U_{i})\negmedspace-\negmedspace H(Y_{i}|X_{i},\negmedspace A_{i},\negmedspace U_{i},\negmedspace V_{i}),\label{eq:R2_end}
\end{align}
where $(a)$ follows since $M_{2}$ is a function of $(M_{1},Y^{n})$,
$(b)$ follows since $A_{i}$ is a function of $(M_{1},Y^{i-1})$
and $(c)$ follows since the Markov chain $Y_{i}\textrm{---}(X_{i},U_{i},A_{i})\textrm{---}X^{i-1}$
holds by the problem definition (this can be easily checked by using
d-separation on the Bayesian network representation of the joint distribution
of the variables at hand as induced by the system model
\begin{figure}[h!]
\centering\includegraphics[bb=177bp 460bp 480bp 703bp,clip,scale=0.8]{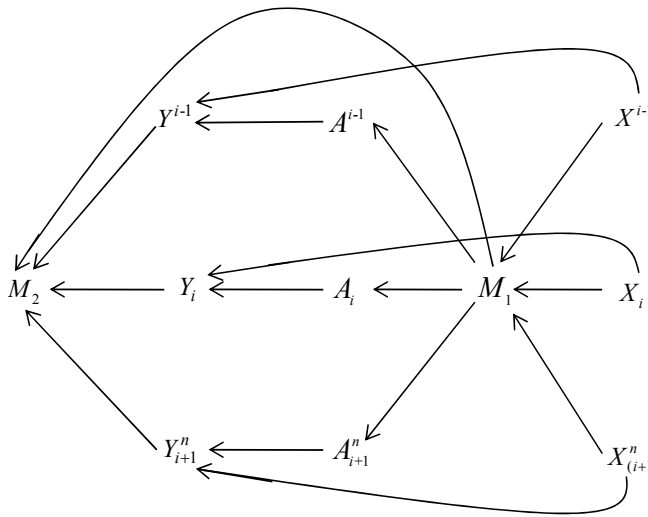}
\caption{Bayesian network representing the joint pmf of variables $(M_{1},M_{2},X^{n},Y^{n},A^{n})$
for the two-way source coding problem with a vending machine in Fig.
\ref{fig:fig1}.}

\label{fig:graph} 
\end{figure}
 in Fig. \ref{fig:graph}, see, e.g., \cite[Sec. A.9]{Kramer}), since
conditioning reduces entropy and by defining $V_{i}=M_{2}$.

Defining $Q$ to be a random variable uniformly distributed over $[1,n]$
and independent of all the other random variables and with $X\overset{\triangle}{=}X_{Q}$,
$Y\overset{\triangle}{=}Y_{Q}$, $A\overset{\triangle}{=}A_{Q}$,
$\hat{X}_{1}\overset{\triangle}{=}\hat{X}_{1Q}$, $\hat{X}_{2}\overset{\triangle}{=}\hat{X}_{2Q}$,
$V\overset{\triangle}{=}(V_{Q},Q)$ and $U\overset{\triangle}{=}(U_{Q},Q),$
from (\ref{eq::R1_end}) we have 
\begin{eqnarray}
nR_{1} & \geq & H(X|Q)-H(X|A,Y,U,Q)\nonumber \\
 &  & +H(Y|X,A,Q)-H(Y|A,Q)\nonumber \\
 & \stackrel{(a)}{\geq} & H(X)-H(X|A,Y,U)\nonumber \\
 &  & +H(Y|X,A)-H(Y|A)\nonumber \\
 & = & I(X;A)+I(X;U|A,Y),
\end{eqnarray}
where $(a)$ follows by the fact that source $X^{n}$ and side information
vending machine are memoryless and since conditioning decreases entropy.
Next, from (\ref{eq:R2_end}), we have
\begin{eqnarray}
nR_{2} & \geq & H(Y|X,A,U)-H(Y|X,A,U,V)\nonumber \\
 & = & I(Y;V|X,A,U).
\end{eqnarray}
Moreover, from Fig. \ref{fig:graph} and using d-separation, it can
be seen that Markov chains $U_{i}\textrm{---}(X_{i},A_{i})\textrm{---}Y_{i}$
and $V_{i}\textrm{---}(A_{i},U_{i},Y_{i})\textrm{---}X_{i}$ hold.
This implies that the random variables $(X,Y,A,U,V)$ factorize as
in (\ref{eq:joint}). 

We now need to show that the estimates $\hat{X}_{1}$ and $\hat{X}_{2}$
can be taken to be functions of $(V,X)$ and $(U,Y)$, respectively.
To this end, recall that, by the problem definition, the reconstruction
$\hat{X}_{1i}$ is a function of $(M_{2},X^{n})$ and thus of $(X_{i},U_{i},V_{i},X^{i-1})$.
Moreover, we can take $\hat{X}_{1i}$ to be a function of $(X_{i},U_{i},V_{i})$
only without loss of optimality, due to the Markov chain relationship
$Y_{i}\textrm{---}(X_{i},U_{i},V_{i})\textrm{---}X^{i-1}$, which
can be again proved by d-separation using Fig. \ref{fig:graph}. This
implies that the distortion $d_{1}(X_{i},Y_{i},X{}_{1}^{i})$ cannot
be reduced by including also $X^{i-1}$ in the functional dependence
of $X{}^{\ensuremath{i}}$. Similarly, the reconstruction $\hat{X}_{2i}$
is a function of $(M_{1},Y^{n})$ by the problem definition, and can
be taken to be a function of $(U_{i},Y_{i})$ only without loss of
optimality, since the Markov chain relationship $X_{i}\textrm{---}(Y_{i},A_{i},U_{i})\textrm{---}Y_{i+1}^{n}$
holds. These arguments and the fact that the definition of $V$ and
$U$ includes the time-sharing variable $Q$ allow us to conclude
that we can take $\hat{X}_{1}$ to be a function of $(U,V,X)$ and
$\hat{X}_{2}$ of $(U,Y)$. We finally observe that $V$ is arbitrarily
correlated with $U$ as per (\ref{eq:joint}) and thus we can without
loss of generality set $\hat{X}_{1}$ to be a function of $(V,X)$
only. The bounds (\ref{eqn: action_const}) follow immediately from
the discussion above and the constraints (\ref{action cost})-(\ref{dist const}). 

To bound the cardinality of auxiliary random variable $U$, we observe
that (\ref{eq:joint}) factorizes as
\begin{equation}
p(x,y,a,u,v)=p(u)p(a,x|u)p(y|a,x)p(v|a,u,y).
\end{equation}
Therefore, for fixed $p(y|a,x),$ $p(a,u|x)$ and $p(v|a,u,y)$ the
characterization in Proposition \ref{prop:RD_action_2way} can be
expressed in terms of integrals $\int g_{j}(\cdot)dF(u),$ for $j=1,...,\left\vert \mathcal{X}\right\vert \left\vert \mathcal{A}\right\vert +3,$
of functions $g_{j}(\cdot)$ of the given fixed pmfs. Specifically,
we have $g_{j}$ for $j=1,...,\left\vert \mathcal{X}_{1}\right\vert \left\vert \mathcal{X}_{2}\right\vert -1,$
given by $p(a,x|u)$ for all values of $x\in\mathcal{X}$ and $a\in\mathcal{A}$
(except one); $g_{\left\vert \mathcal{X}_{1}\right\vert \left\vert \mathcal{X}_{2}\right\vert }=H(X|A,Y,U=u)$;
$g_{\left\vert \mathcal{X}_{1}\right\vert \left\vert \mathcal{X}_{2}\right\vert +1}=I(Y;V|A,X,U=u);$
$g_{\left\vert \mathcal{X}_{1}\right\vert \left\vert \mathcal{X}_{2}\right\vert +2}=\mathrm{E}[d_{1}(X,Y,\ce{f}_{1}(V,X))|U=u]$
and $g_{\left\vert \mathcal{X}_{1}\right\vert \left\vert \mathcal{X}_{2}\right\vert +3}=\mathrm{E}[d_{2}(X,Y,\ce{f}_{2}(U,Y))|U=u]$.
The proof is concluded by invoking Caratheodory Theorem. 

To bound the cardinality of auxiliary random variable $V,$ we note
that (\ref{eq:joint}) can be factorized as
\begin{equation}
p(x,y,a,u,v)=p(v)p(a,y,u|v)p(x|a,u,y),
\end{equation}
so that, for fixed $p(x|a,u,y)$, the characterization in Proposition
\ref{prop:RD_action_2way} can be expressed in terms of integrals
$\int g_{j}(p(a,u,y|v))dF(v),$ for $j=1,...,|\mathcal{A}||\mathcal{U}||\mathcal{Y}|+1,$
of functions $g_{j}(\cdot)$ that are continuous on the space of probabilities
over alphabet $|\mathcal{A}|\times|\mathcal{U}|\times\left\vert \mathcal{Y}\right\vert .$
Specifically, we have $g_{j}$ for $j=1,...,|\mathcal{A}||\mathcal{U}||\mathcal{Y}|-1,$
given by $p(a,u,y)$ for all values of $a\in\mathcal{A}$, $u\in\mathcal{U}$
and $y\in\mathcal{Y}$ (except one); $g_{|\mathcal{A}||\mathcal{U}||\mathcal{Y}|}=H(Y|A,X,U,V=v);$
and $g_{|\mathcal{A}||\mathcal{U}||\mathcal{Y}|+1}=\mathrm{E}[d_{1}(X,Y,\ce{f}_{1}(V,X))|V=v]$.
The proof is concluded by invoking Fenchel--Eggleston--Caratheodory
Theorem \cite[Appendix C]{Elgammal}.

The converse for Proposition \ref{prop:RD_action_2way_ind} follows
similar steps as above with the only difference that here we have
\begin{align}
 & nR_{1}\stackrel{(a)}{\geq}\sum_{i=1}^{n}H(Z_{i})-H(Z_{i}|Z_{i+1}^{n},Y^{n},M_{1},A^{i})\nonumber \\
 & +H(Y_{i}|Y^{i-1},Z^{n},M_{1},A_{i})-H(Y_{i}|Y^{i-1},M_{1},A_{i})\nonumber \\
 & \stackrel{(b)}{\geq}\sum_{i=1}^{n}H(Z_{i})\negmedspace-\negmedspace H(Z_{i}|A_{i},\negmedspace Y_{i},\negmedspace U_{i})\negmedspace\nonumber \\
 & +\negmedspace H(Y_{i}|Z_{i},\negmedspace A_{i})\negmedspace-\negmedspace H(Y_{i}|A_{i}),\label{eq:R1_end_ind}
\end{align}
where $(a)$ follows follows as in ($a$)-($b$) of (\ref{eq::R1_end});
and ($b$) follows since Markov chain relationship $Y_{i}\textrm{---}(Z_{i},A_{i})\textrm{---}(Y^{i-1},Z^{n\backslash i},M_{1})$
holds. The rest of the proof is as above.

\section*{Appendix B: Proofs for the Example in Sec. \ref{sub:Example}}

1) $D_{1}=D_{1,max}$ and $D_{2}=0$: Here, we prove that the rate-cost
region in Proposition \ref{prop:RD_action_2way_ind} is given by (\ref{eqn: r_ex1})
for $D_{1}=D_{1,max}$ and $D_{2}=0$. We begin with the converse
part. Starting from (\ref{eq:R1_ind}), we have
\begin{eqnarray}
R_{1} & \stackrel{(a)}{\geq} & I(A;Z)+H(Z|A,Y)\nonumber \\
 & = & H(Z)-I(Z;Y|A)\nonumber \\
 & \stackrel{(b)}{\geq} & H(Z)-\Gamma I(Z;X|A=1)\label{eq:ex1_convR1_firststeps}\\
 & \stackrel{(c)}{\geq} & H(Z)-\Gamma H(X|A=1)\nonumber \\
 & \stackrel{(d)}{\geq} & H(Z)-\Gamma\nonumber \\
 & \stackrel{(e)}{=} & H_{2}(\epsilon)+1-\epsilon-\Gamma,
\end{eqnarray}
where ($a$) follows from (\ref{eq:R1_ind}) and since $Z$ has to
be recovered losslessly at Node 2; ($b$) follows since $\ce{Pr}[A=1]=\ce{E}[\Lambda(A)]\leq\Gamma$;
($c$) follows because entropy is non-negative; ($d$) follows since
$H(X|A=1)\leq1$; and ($e$) follow because $H(Z)=H_{2}(\epsilon)+1-\epsilon$.
Achievability follows by setting $U=Z$, $V=\emptyset$, $\ce{Pr}(A=1|Z=0)=\ce{Pr}(A=1|Z=1)=\nicefrac{\Gamma}{(1-\epsilon)}$
and $\ce{Pr}(A=0|Z=\ce{e})=1$ in (\ref{eqn: RD_action_2way_ind}).

2) $D_{1}=0$ and $D_{2}=D_{2,max}$: Here, we turn to the case $D_{1}=0$
and $D_{2}=D_{2,max}$. We start with the converse. Since $X$ is
to be reconstructed losslessly at Node 1, we have the requirement
$H(X|V,Z)=0$ from (\ref{eq:dist1_ind}). It easy to see that this
requires that the equalities $A=1$ and $V=Y=X$ be met if $Z=\ce{e}$.
In fact, otherwise, $X$ could not be a function of $(V,Z)$ as required
by the equality $H(X|V,Z)=0$. The condition that $A=1$ if $Z=\ce{e}$
requires that the pmf $p(a|z)$ be such that $\ce{Pr}(A=1|Z=\ce{e})=1$,
which entails $\Gamma=\ce{Pr}[A=1]\geq\ce{Pr}[Z=\ce{e}]=\epsilon$.
Moreover, we can set $\ce{Pr}(A=1|Z=0)=\ce{Pr}(A=1|Z=1)=\nicefrac{(\gamma-\epsilon)}{(1-\epsilon)}$,
for some $0\leq\gamma\leq\Gamma$, by leveraging the symmetry of the
problem on the selection of the actions given $Z=0$ and $Z=1$. Starting
from (\ref{eq:R1_ind}), we can thus write
\begin{eqnarray}
R_{1} & \stackrel{(a)}{\geq} & I(Z;A)\nonumber \\
 & = & H(Z)-H(Z|A)\nonumber \\
 & = & H_{2}(\epsilon)+1-\epsilon-\gamma H(Z|A=1)\nonumber \\
 & - & (1-\gamma)H(Z|A=0)\nonumber \\
 & \stackrel{(b)}{=} & H_{2}(\epsilon)+1-\epsilon-\gamma H\Bigl(\frac{\epsilon}{\gamma},\frac{\gamma-\epsilon}{2\gamma},\frac{\gamma-\epsilon}{2\gamma}\Bigr)\nonumber \\
 &  & -(1-\gamma)\nonumber \\
 & = & H_{2}(\epsilon)-\gamma H_{2}\bigl(\frac{\epsilon}{\gamma}\bigr)\nonumber \\
 & \stackrel{(a)}{\geq} & H_{2}(\epsilon)-\Gamma H_{2}\bigl(\frac{\epsilon}{\Gamma}\bigr),
\end{eqnarray}
where ($a$) follows from (\ref{eq:R1_ind}) and since there is no
distortion requirement at Node 2; ($b$) follows by direct calculation;
and ($c$) follows since $H_{2}(\epsilon)-\gamma H_{2}(\frac{\epsilon}{\gamma})$
is minimzed at $\gamma=\Gamma$ over all $0\leq\gamma\leq\Gamma$.

The bound (\ref{eq:R2_ex2}) follows immediately by providing Node
2 with the sequence $X^{n}$ and then using the bound $R_{2}\geq H(X|Z)=\epsilon$. 

Achievability follows by setting $U=\emptyset$ and the pmf $p(a|z)$
be such that $\ce{Pr}(A=1|Z=\ce{e})=1$ and $\ce{Pr}(A=1|Z=0)=\ce{Pr}(A=1|Z=1)=\frac{\Gamma-\epsilon}{1-\epsilon}$.
Moreover, let $V=Y=X$ if $Z=\ce{e}$ and $V=Y=\phi$ otherwise. Evaluating
(\ref{eqn: RD_action_2way_ind}) with these choices leads to (\ref{eqn: r_ex2}). 

3) $D_{1}=D_{2}=0$: Here, we prove the rate-cost region (\ref{eqn: r_ex3})
for the case $D_{1}=D_{2}=0$. Starting from (\ref{eq:R1_ind}), we
have
\begin{eqnarray}
R_{1} & \stackrel{(a)}{\geq} & H(Z)-\Gamma I(Z;X|A=1)\nonumber \\
 & \stackrel{(b)}{=} & H(Z)-\Gamma H(X|A=1)\nonumber \\
 &  & +\Gamma H(X|A=1,Z=\ce{e})\ce{Pr}(Z=\ce{e}|A=1)\nonumber \\
 & \stackrel{(c)}{\geq} & H(Z)-\Gamma+\Gamma.\frac{\epsilon}{\Gamma}\nonumber \\
 & \stackrel{}{=} & H_{2}(\epsilon)+1-\Gamma,
\end{eqnarray}
where ($a$) follows as in (\ref{eq:ex1_convR1_firststeps}); ($b$)
follows because $H(X|A=1,Z=0)=H(X|A=1,Z=1)=0$; ($c$) follows since
$H(X|A=1)\leq1$, $H(X|A=1,Z=\ce{e})=1$ and because $p(Z=\ce{e}|A=1)=\frac{\epsilon}{\Gamma}$,
where latter follows from the the requirement $H(X|V,Z)=0$ as per
discussion provided in the previous section.

For the achievability, let $U=Z$, $\ce{Pr}(A=1|Z=\ce{e})=1$ and
$\ce{Pr}(A=1|Z=0)=\ce{Pr}(A=1|Z=1)=\frac{\Gamma-\epsilon}{1-\epsilon}$.
Moreover, let $V=Y=X$ if $Z=\ce{e}$ and $V=Y=\emptyset$ otherwise.
Evaluating (\ref{eqn: RD_action_2way_ind}) with these choices leads
to (\ref{eqn: r_ex3}).

\section*{Appendix C: Converse Proof for Proposition \ref{prop:RD_action_2way-HB_ind}}

Here, we prove the converse part of Proposition \ref{prop:RD_action_2way-HB_ind}.
For any $(n,R_{1},R_{2},D_{1}+\epsilon,D_{2}+\epsilon,D_{3}+\epsilon,\Gamma+\epsilon)$
code, we have the series of inequalities
\begin{align}
 & nR_{1}\geq H(M_{1})\nonumber \\
 & \stackrel{(a)}{\geq}\sum_{i=1}^{n}H(Z_{i})-H(Z_{i}|Z_{i+1}^{n},Y^{n},M_{1},A^{i},\hat{X}_{3i})\nonumber \\
 & +\negmedspace H(Y_{i}|Y^{i-1},\negmedspace X^{n},\negmedspace M_{1},\negmedspace A_{i},\negmedspace\hat{X}_{3i})\negmedspace-\negmedspace H(Y_{i}|Y^{i-1},\negmedspace M_{1},\negmedspace A_{i},\negmedspace\hat{X}_{3i})\nonumber \\
 & \stackrel{(b)}{\geq}\sum_{i=1}^{n}\negmedspace H(Z_{i})\negmedspace-\negmedspace H(Z_{i}|A_{i},\negmedspace Y_{i},\negmedspace U_{i},\negmedspace\hat{X}_{3i})\negmedspace+\negmedspace H(Y_{i}|Z_{i},\negmedspace A_{i},\negmedspace\hat{X}_{3i})\nonumber \\
 & -H(Y_{i}|A_{i},\hat{X}_{3i}),\label{eq:R1_end_HB}
\end{align}
where $(a)$ follows from ($a$) in (\ref{eq:R1_end_ind}) by noting
that $\hat{X}_{3i}$ is a function of $M$ and $(b)$ follows since
conditioning decreases entropy, by defining $U_{i}=(M_{1},X_{i+1}^{n},A^{i-1},Y^{i-1})$
and using the Markov chain relationship $Y_{i}\textrm{---}(Z_{i},A_{i},\hat{X}_{3i})\textrm{---}(Y^{i-1},X^{n\backslash i},M_{1})$.
We also have the series of inequalities
\begin{align}
 & nR_{2}\geq H(M_{2})\nonumber \\
 & \stackrel{(a)}{\geq}\sum_{i=1}^{n}H(Y_{i}|Z_{i},A_{i},U_{i},\hat{X}_{3i})\negmedspace-\negmedspace H(Y_{i}|Z_{i},A_{i},U_{i},V_{i},\hat{X}_{3i}),\label{eq:R2_end_HB}
\end{align}
where ($a$) follows from (\ref{eq:R2_end}), by replacing sequence
$X^{n}$ with the sequence $Z^{n}$ and by observing that $\hat{X}_{3i}$
is a function of $M_{1}$. Defining $Q$ as in Appendix A, along with
$\hat{X}_{3}\overset{\triangle}{=}\hat{X}_{3Q}$, from (\ref{eq:R1_end_HB})
we have
\begin{align*}
nR_{1} & \geq H(Z|Q)-H(Z|A,Y,U,\hat{X}_{3},Q)\\
 & +H(Y|Z,A,\hat{X}_{3},Q)-H(Y|A,\hat{X}_{3},Q)\\
 & \stackrel{(a)}{\geq}H(Z)-H(Z|A,Y,U,\hat{X}_{3})\\
 & +H(Y|Z,A,\hat{X}_{3})-H(Y|A,\hat{X}_{3})\\
 & =\negmedspace I(Z;A)\negmedspace+\negmedspace I(Z;\hat{X}_{3}|A)\negmedspace+\negmedspace I(Z;U|A,\negmedspace Y,\negmedspace\hat{X}_{3}),
\end{align*}
where $(a)$ follows by the fact that source $Z^{n}$ and side information
vending machine are memoryless and since conditioning decreases entropy.
Next, from (\ref{eq:R2_end_HB}), we have
\begin{eqnarray}
nR_{2} & \geq & H(Y|Z,A,U,\hat{X}_{3})-H(Y|Z,A,U,V,\hat{X}_{3})\nonumber \\
 & = & I(Y;V|Z,A,U,\hat{X}_{3}).
\end{eqnarray}
Moreover, by just adding $\hat{X}_{3}^{n}$ to the Bayesian graph
in Fig. \ref{fig:graph}and using d-separation, it can be seen that
Markov chains $U_{i}\textrm{---}(Z_{i},A_{i})\textrm{---}Y_{i}$ and
$V_{i}\textrm{---}(A_{i},U_{i},Y_{i},\hat{X}_{3})\textrm{---}Z_{i}$
hold, which implies that the random variables $(X,Y,Z,A,U,V,\hat{X}_{3})$
factorize as in (\ref{eq:joint-HB_ind}). Based on the discussion
in the converse proof in Appendix A, it is easy to see that the estimates
$\hat{X}_{1}$ and $\hat{X}_{2}$ are functions of $(V,X)$ and $(U,Y)$,
respectively. The bounds (\ref{eqn: RD_action_2way-HB_ind}) follow
immediately from the discussion above and the constraints (\ref{action cost})-(\ref{dist const})
and (\ref{dist const-HB}).

\end{document}